\documentclass[doublecol]{epl2} 

\usepackage{amsmath,amssymb}
\usepackage{epstopdf}
\usepackage{bm}
\title{Effects of an embedding bulk fluid on phase separation dynamics in a thin liquid film}
\shorttitle{Effects of an embedding bulk fluid on phase separation dynamics in a thin liquid film} 

\author{
S. Ramachandran\inst{1},
S. Komura\inst{1}\thanks{E-mail: \email{komura@tmu.ac.jp}} 
\and 
G. Gompper\inst{2}
}

\shortauthor{S. Ramachandran \etal}

\institute{                    
\inst{1} 
Department of Chemistry, 
Graduate School of Science and Engineering, 
Tokyo Metropolitan University, 
Tokyo 192-0397, Japan.\\
\inst{2}
Institut f\"ur Festk\"orperforschung, 
Forschungszentrum J\"ulich - D-52425 J\"ulich, Germany, EU
}


\abstract{
Using dissipative particle dynamics simulations, we study the effects 
of an embedding bulk fluid on the phase separation dynamics in a  
thin planar liquid film.
The domain growth exponent is altered from 2D to 3D behavior 
upon the addition of a bulk fluid, even though the phase separation occurs 
in 2D geometry.
Correlated diffusion measurements 
in the film show that the 
presence of bulk fluid changes the nature of the longitudinal 
coupling diffusion coefficient from logarithmic to algebraic 
dependence of $1/s$, where $s$ is the distance between the 
two particles.
This result, along with the scaling exponents, suggests that the phase 
separation takes place through the Brownian coagulation process.
}

\begin{document}

\maketitle

\section{Introduction}
\label{intro}

Lipid molecules constituting the 
membranes of biological cells play a major role in the regulation 
of various cellular processes.
About a decade ago Simons and Ikonen proposed a hypothesis
which suggests that the lipids organize themselves into 
sub-micron sized domains termed ``rafts"~\cite{simons-97}.
The rafts serve as platforms for proteins, which  in turn attributes
a certain functionality to each domain.
Although there have been extensive studies in this area, the details of the 
underlying physical mechanisms leading to formation of rafts, their stability, 
and the regulation of their finite domain size remain elusive.
Numerous experiments on intact cells and artificial membranes containing 
mixtures of saturated lipids, unsaturated lipids and cholesterol, 
have demonstrated the segregation of the lipids into liquid-ordered 
and liquid-disordered phases~\cite{veatch-05}.
Recent experimental observations of critical fluctuations  
point towards the idea that the cell maintains its membrane slightly 
above the critical point~\cite{hsmith-08,veatch-08}.  
Below the transition temperature, there have been studies on the 
dynamics in multicomponent membranes such as diffusion of 
domains~\cite{cicuta-07} and domain coarsening~\cite{yanagisawa-07}.  
A clear understanding of phase separation may contribute 
towards a better explanation of the dynamics of lipid organization 
in cell membranes.
Apart from biological membranes, it is of relevance to understand
the dynamics of Langmuir monolayer systems which are also thin fluid 
films.

Phase separation of binary fluids following a quench has been under 
study for over forty years~\cite{bray-02}.
The dynamic scaling hypothesis assumes that there exists a scaling 
regime characterized by the average domain size $R$ that grows with
time $t$ as $R \sim t^\alpha$ with an universal exponent $\alpha$.
For three-dimensional (3D) off-critical binary fluids, there is an 
initial growth by the Brownian coagulation process~\cite{binder-74}, 
followed by the Lifshitz-Slyozov (LS) evaporation-condensation
process~\cite{lifshitz-81}; both mechanisms show a growth exponent 
$\alpha=1/3$.  
This is followed by a late time inertial regime of 
$\alpha=2/3$~\cite{furukawa-94}. 
For critical mixtures, there is an intermediate $\alpha=1$ regime 
owing to interface diffusion~\cite{siggia-79}.
The scenario is slightly different for pure two-dimensional (2D) 
systems~\cite{miguel-85}.         
For an off-critical mixture, it was predicted that after the initial 
formation of domains, they grow by the Brownian coagulation mechanism
with a different exponent $\alpha=1/2$ (as will be explained later),
followed by a crossover to the LS mechanism which gives $\alpha=1/3$ 
even in 2D.
For critical mixtures, on the other hand, the initial quench produces 
an interconnected structure which coarsens and then breaks up due 
to the interface diffusion with an exponent $\alpha=1/2$.
After the breakup processes, coarsening takes place through Brownian
coagulation that is again characterized by the $\alpha=1/2$ 
scaling~\cite{binder-74}.
These predictions were confirmed by molecular dynamics simulations
in 2D~\cite{ossadnik-94}.
The exponent $\alpha=1/2$ was also observed in 2D lattice-Boltzmann 
simulations in the presence of thermal noise for a critical 
mixture~\cite{yeomans-99}.

Although biomembranes composed of lipid bilayers can be regarded as 
2D viscous fluids, they are not isolated pure 2D systems since 
lipids are coupled to the adjacent fluid.
Hence it is of great interest to investigate the phase separation 
dynamics in such a quasi-2D liquid film in the presence of hydrodynamic 
interaction.   
(We use the word ``quasi-2D'' whenever the film is coupled to 
the bulk fluid.)
To address this problem, we consider a 2D binary viscous fluid 
in contact with a bulk fluid.  
Our approximation of the membrane as a planar 2D liquid film is
valid in the limit of large bending rigidity (common in biological
membranes) or in the presence of a lateral tension, which both 
act to suppress membrane undulations.
We employ a simple model in which the film is confined 
to a plane with the bulk fluid particles added above and below. 
In our model using dissipative particle dynamics (DPD) simulation 
technique, the exchange of momentum between the film
 and the bulk fluid is naturally taken into account.  
We particularly focus on the effect of bulk fluid on the quasi-2D 
phase separation. 
We show that the presence of a bulk fluid will alter the domain growth 
exponent from that of 2D to 3D indicating the significant role 
played by the film-solvent coupling.   
In order to elucidate the underlying physical mechanism of this effect,
we have looked into the diffusion properties in the film
by measuring two-particle correlated diffusion.        
Our result suggests that quasi-2D phase separation proceeds by the 
Brownian coagulation mechanism which reflects the 3D nature of the bulk 
fluid. 
Such a behavior is universal as long as the domain size exceeds the 
Saffman-Delbr\"uck length~\cite{saffman-76}.

\section{Model and simulation technique}
\label{model}

\begin{figure}[t]
\begin{center}
\includegraphics[scale=0.2]{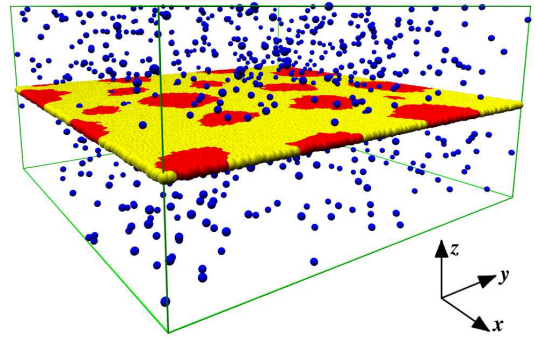}
\caption{Image of the model film with the bulk fluid called 
solvent.
The yellow and red particles represent the two components constituting
the model film, while blue ones represent the solvent.
For clarity, only a fraction of the solvent particles are shown.}
\label{image}
\end{center}
\end{figure}

For the purpose of our study, we use a structureless 
model of the 2D liquid film within  
the DPD framework~\cite{degroot-warren-97,espanol-warren-95}.
As shown in fig.~\ref{image}, the 2D film is represented by 
a single layer of particles confined to a  plane.
In order to study phase separation, we introduce two species 
of particles, $A$ and $B$. 
The bulk fluid which we call as ``solvent'' ($S$) is also represented by 
single particles of same size as that of the film particles.  
All particles have the  same mass $m$.  
We avoid using the existing DPD models for 
a self-assembling bilayer~\cite{sunil-mohamed-05, sanoop-08} as they inherently
include bending and protrusion modes, which makes it
difficult to separate hydrodynamic effects from the effect
of membrane shape deformations.

In DPD, the interaction between any two particles, within a range $r_0$, 
is linearly repulsive. 
The pairwise interaction leads to full momentum conservation, 
which in turn brings out the correct fluid hydrodynamics.
The force on a particle $i$ is given by 
\begin{equation}
m\frac{\upd \mathbf{v}_i}{\upd t}  
= \sum_{j\neq i} \left[ 
\mathbf{F}_{ij}^{\rm C}(\mathbf{r}_{ij}) +
\mathbf{F}_{ij}^{\rm D}(\mathbf{r}_{ij},\mathbf{v}_{ij}) +
\mathbf{F}_{ij}^{\rm R}(\mathbf{r}_{ij})
\right],
\label{eqn:EoM}
\end{equation} 
where $\mathbf{r}_{ij} = \mathbf{r}_i - \mathbf{r}_j$ 
and $\mathbf{v}_{ij} = \mathbf{v}_i - \mathbf{v}_j$.
Of the three types of forces acting on the particles, 
the conservative force on particle $i$ due to $j$ is 
$\mathbf{F}_{ij}^{\rm C}=
a_{ij}\omega(r_{ij})\hat{\mathbf{r}}_{ij}$,
where $a_{ij}$ is an interaction strength and 
$\hat{\mathbf{r}}_{ij}=\mathbf{r}_{ij}/r_{ij}$ with 
$r_{ij}=|\mathbf{r}_{ij}|$.
The second type of force is the dissipative force 
$\mathbf{F}_{ij}^{\rm D}=  
-\Gamma_{ij}\omega^2(r_{ij})
(\hat{\mathbf{ r}}_{ij}\cdot{\bf v}_{ij})\hat{\mathbf{r}}_{ij}$,
where $\Gamma_{ij}$ is the dissipative strength for the pair $(i,j)$.
The last is the random force
$\mathbf{F}_{ij}^{\rm R}= 
\sigma_{ij}(\Delta t)^{-1/2}\omega(r_{ij})
\zeta_{ij}\hat{\mathbf{r}}_{ij}$,
where $\sigma_{ij}$ is the amplitude of the random noise 
for the pair $(i,j)$, and $\zeta_{ij}$ is a random variable 
with zero mean and unit variance which is uncorrelated for 
different pairs of particles and different time steps.
The dissipative and random forces act as a thermostat, provided the 
fluctuation-dissipation theorem $\sigma_{ij}^2=2\Gamma_{ij} k_{\rm B}T$ 
is satisfied ($k_{\rm B}$ is Boltzmann constant and $T$ is the 
thermostat temperature).
The weight factor is chosen as $\omega(r_{ij})= 1-r_{ij}/r_0$ up to 
the cutoff radius $r_0$ and zero thereafter.
The particle trajectories are obtained by solving eq.~(\ref{eqn:EoM})
using the velocity-Verlet integrator. 
In the simulation, $r_0$ and $m$ set the scales for length and mass,  
respectively, while $k_{\rm B}T$ sets the energy scale. 
The time is measured in units of 
$\tau=\left(mr_0^2/k_{\rm B}T\right)^{1/2}$. 
The numerical value of the amplitude of the random force is assumed 
to be the same for all pairs such that 
$\sigma_{ij}=3.0\left[(k_{\rm B}T)^3 m/r_0^2 \right]^{1/4}$,
and the fluid density is set as $\rho=3.0$. 
We set $k_{\rm B}T=1$ and the integration time step is 
chosen to be $\Delta t=0.01\tau$.

\begin{figure}[t]
\begin{center}
\includegraphics[scale=0.3]{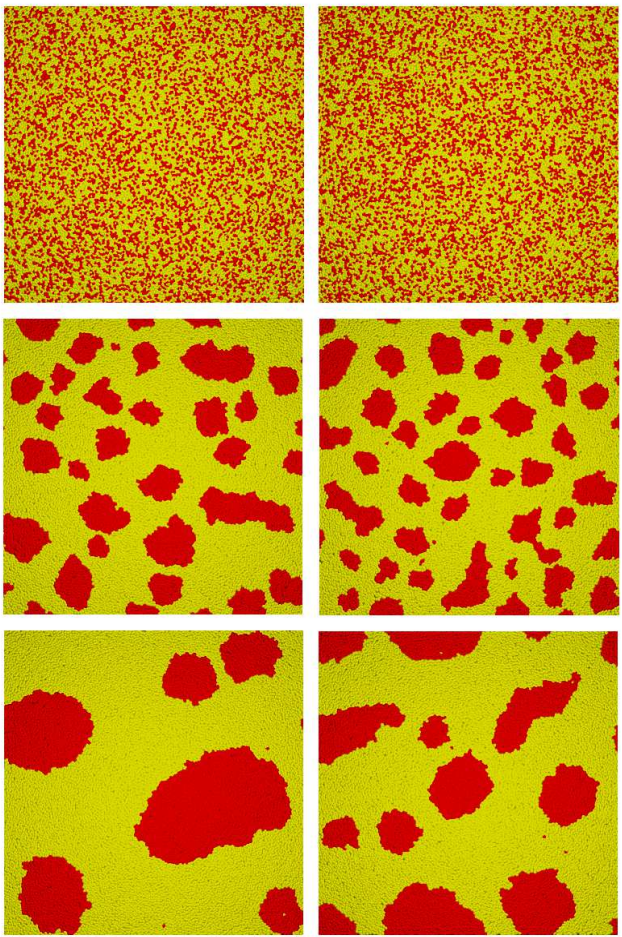}
\caption{The snapshots for a 70:30 mixture (yellow and red) undergoing 
phase separation at $t=0,150$ and $1000$ (top to bottom) for a pure 2D 
(left column) and quasi-2D system with $L_z=40$ (right column).
The above snapshots are from one of the ten independent trials that were
conducted.
}
\label{fig:panel7030}
\end{center}
\end{figure}

The thin film is constructed by placing particles in the $xy$-plane 
in the middle of the simulation box (see fig.~\ref{image}). 
Owing to the structureless representation of the constituent particles, we
apply an external potential so as to maintain the film integrity.
This is done by fixing the $z$-coordinates of all the film particles. 
It is known that confinement of simulation particles between walls
leads to a reduction in the solvent temperature near the 
wall~\cite{altenhoff-07}.
However, since we allow for the in-plane motion of the film particles,
the solvent temperature is found to be only 2\% less than the bulk near 
the 2D film.
Hence we consider that this effect is negligible.
Our work involves the systematic variation of the height of the
simulation box starting from the pure 2D case.
In the absence of solvent, we work with a 2D-box of dimensions 
$L_x \times L_y = 80\times 80$ with $19,200$ particles constituting 
the film.   
For the quasi-2D studies, we add solvent particles $S$ above and below 
the model film, and increase the height of the box as 
$L_z = 5, 20$ and $40$.  
For all the cases there are $19,200$ film particles.
The largest box size ($L_z=40$) has $748,800$ solvent particles.
The box with height $L_z= 40$ is found to be sufficiently large enough 
to prevent the finite size effect which affects the solvent-film 
interaction.  
The system is then subject to periodic boundary conditions in all the three 
directions. 
For phase separation simulations, we introduce two species of 
 film particles $A$ and $B$.
The interaction parameter between various particles are given by
$a_{AA}=a_{BB}=a_{SS}=a_{AS}=a_{BS}=25$ and $a_{AB}=50$.
In order to do a quench, the film is first equilibrated with
a single component, following which a fraction of the particles are
instantaneously changed to the $B$ type.

\section{Phase separation} 
\label{separation}

\begin{figure}[t]
\begin{center}
\includegraphics[scale=0.4]{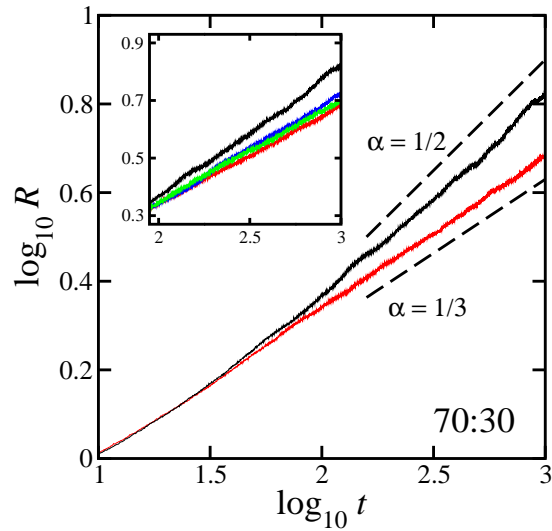}
\caption{The average domain size $R$ as a function of time $t$ 
for a $70:30$ off-critical mixture.  
The upper black curve is the pure 2D case showing an $\alpha=1/2$ 
scaling, and the lower red curve is the quasi-2D case when $L_z = 40$
showing a distinct $\alpha=1/3$ scaling.
The inset shows the zoomed in portion with different box heights
in the $z$-direction, i.e., $L_z = 0, 5, 20$ and $40$ starting 
from the top black curve.
}
\label{fig:7030}
\end{center}
\end{figure}

First we describe the results of the phase separation dynamics. 
The snapshots for $A:B$ composition set to $70:30$ (off-critical mixture)
is shown in fig.~\ref{fig:panel7030} for both pure 2D case 
(left column) and quasi-2D case with $L_z=40$ (right column). 
Qualitatively, it is seen that the domains for the quasi-2D case 
are smaller in size when compared at the same time step.
We also monitor the average domain size $R(t)$ which can be obtained 
from the total interface length $L(t)$ between the two components.
This is because $R(t)$ and $L(t)$ are related by $L(t)=2 \pi N(t) R(t)$,
where $N(t)$ is the number of domains. 
The area occupied by the $B$-component is given by 
${\cal A}=\pi N(t) R^2(t)$ which is a conserved quantity.
Then we have 
\begin{equation}
R(t) = 2{\cal A}/L(t).
\end{equation}
When the domain size grows as $R \sim t^{\alpha}$, one has
$L \sim t^{-\alpha}$ and $N \sim t^{-2\alpha}$.
The domain size $R(t)$ for $70:30$ mixture is shown in fig.~\ref{fig:7030}.
In this plot, average over 10 independent trials has been taken.     
It can be seen that the pure 2D case has a growth exponent $\alpha=1/2$.
Upon the addition of solvent, we observe that the exponent shifts 
to a lower value of $\alpha=1/3$.
This exponent is reminiscent of the phase separation dynamics of an 
off-critical mixture in 3D.
Upon systematically increasing the amount of solvent in the system by 
changing the height $L_z$, we can see a clear deviation from the 
pure 2D behavior (see the inset of fig.~\ref{fig:7030}).
There is no further change if $L_z$ is increased beyond $40$.
We note that the scaling regime covers about one decade
in time, which is similar to that previously shown in 
the literature~\cite{sunil-mohamed-05}. 
A larger system size 
$L_x \times L_y = 200 \times 200$
 also produced the same scaling for the pure 2D 
case, which demonstrates that finite-size effects are small. 
However, we present here only the results for the $80\times80$ 
system in 2D, because this is the system size studied for 
the quasi-2D case with a bulk fluid, which requires 
a large amount of particles in 3D.

\begin{figure}[t]
\begin{center}
\includegraphics[scale=0.4]{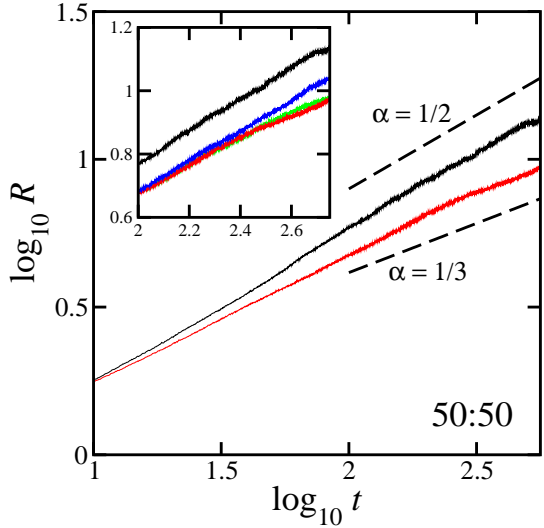}
\caption{The average domain size $R$ as a function of time $t$ for a 
$50:50$ critical mixture.  
The upper black curve is the pure 2D case showing an $\alpha=1/2$ 
scaling, and the lower red curve is the quasi-2D case when $L_z = 40$
showing a distinct $\alpha=1/3$ scaling.
The inset shows the zoomed in portion with different box heights
in the $z$-direction as in fig.~\ref{fig:7030}.
}
\label{fig:5050}
\end{center}
\end{figure}

In fig.~\ref{fig:5050}, we show the result for a component ratio of 
$50:50$ (critical mixture). 
In this case, the growth exponent for the pure 2D case is less obvious
owing to rapid coarsening of the domains.
However, by simulating a bigger system $200\times200$
with the same areal density, an $\alpha=1/2$ exponent is indeed obtained.
Similar to the off-critical case, the growth of the domains is slowed 
down by the addition of solvent and the exponent is reduced to 
$\alpha=1/3$.
These results indicate that solvent is responsible for slowing down 
the growth dynamics.

The observed exponent $\alpha=1/2$ in pure 2D systems can be explained
in terms of the Brownian coagulation mechanism~\cite{miguel-85}.
From dimensional analysis, the 2D diffusion coefficient of the domain 
is given by $D_{2} \sim k_{\rm B}T/\eta$, where $\eta$ is the  
film 2D viscosity.
Using the relation 
\begin{equation}
R^2 \sim D_{2} t \sim (k_{\rm B}T/\eta) t, 
\end{equation} 
we find $R \sim t^{1/2}$. 
For 3D systems, on the other hand, the diffusion coefficient of the 
droplet is inversely proportional to its size, $D_{3} \sim 1/R$, 
a well-known Stokes-Einstein relation. 
Hence the Brownian coagulation mechanism in 3D gives rise to an 
exponent $\alpha=1/3$.
(In general, the exponent is $\alpha=1/d$, where $d$ is the space dimension.)
The change in the exponent from $\alpha=1/2$ to $1/3$ due to the 
addition of solvent implies the crossover from 2D to 3D behaviors of the
phase separation dynamics even though the lateral coarsening takes place 
only within the 2D geometry~\cite{LS}.
Therefore it is necessary to examine the size dependence of the 
domain diffusion coefficient in quasi-2D systems.   
This can be calculated by 
tracking the mean-squared displacement of domains of various radii.  
The equivalent information can be more efficiently obtained by 
calculating the two-particle longitudinal coupling diffusion 
coefficient in a single component film rather than in a binary 
 system.
This is described in the next section.

\section{Correlated diffusion} 
\label{diffusion}

Consider a pair of particles separated by a 2D vector $\mathbf{s}$,
undergoing diffusion in the liquid film.
The two-particle mean squared displacement is given 
by~\cite{diamant-09} 
\begin{equation}
\langle \Delta s_i^k \Delta s_j^l \rangle 
= 2 D_{ij}^{kl}(\mathbf{s})t,
\end{equation}
where $\Delta s_i^k$ is the displacement of the particle $k (=1,2)$ 
along the axis $i (=x,y)$, $D_{ij}^{kl}$ is the diffusion tensor giving 
self-diffusion when $k=l$ and the coupling between them when $k \neq l$.
The $x$-axis is defined along the line connecting a pair of 
particles $1$ and $2$, 
i.e., $\mathbf{s}= s \hat{x}$. 
Hence, we have $D_{xy}^{12}=0$ by symmetry.
The longitudinal coupling diffusion coefficient, 
$D_{\rm L}^{\rm c}(s)=D_{xx}^{12}(s \hat{x})$,
gives the coupled diffusion along the line of centers of the particles.
We first describe the analytical expression of $D_{\rm L}^{\rm c}(s)$ 
based on the Saffman and Delbr\"uck (SD) theory which was originally 
developed for protein diffusion in membranes~\cite{saffman-76}.

\begin{figure}[t]
\begin{center}
\includegraphics[scale=0.4]{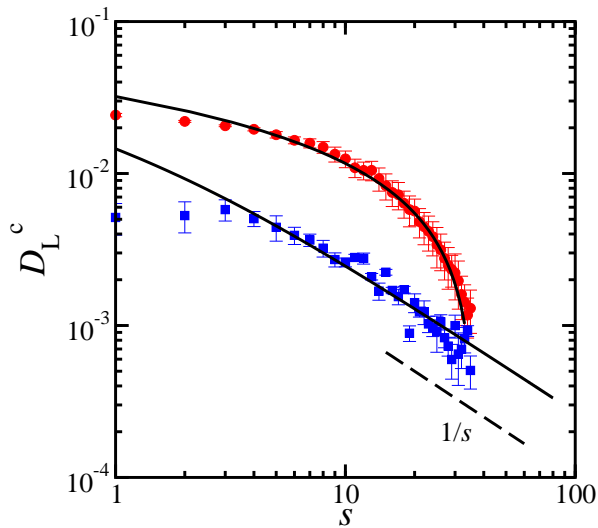}
\caption{
Longitudinal coupling diffusion $D_{\rm L}^{\rm c}$ as a function of 
particle separation $s$.  
The upper circles are data for the pure 2D case.
The lower squares correspond to the case with solvent when $L_z=40$.
The upper solid line is the fit by eq.~(\ref{eqn:longDsmallr}), and the 
lower solid line is the fit by eq.~(\ref{eqn:longD}).
The dashed line shows the $1/s$ dependence.
}
\label{fig:2pcorr}
\end{center}
\end{figure}

Since the calculation of the diffusion coefficient in a pure 2D system
is intractable due to Stokes paradox, SD circumvented this problem by 
taking into account the presence of the solvent with 3D viscosity 
$\eta_{\rm s}$ above and below the membrane.  
Suppose a point force $\mathbf{f}$ directed along the plane of the 
 film lying in the $xy$-plane acts at the origin.
Then we seek for the velocity $\mathbf{v(s)}$ induced at the position 
$\mathbf{s}$. 
According to the SD theory, it is given in 
Fourier space, 
$\mathbf{v}[\mathbf{q}] = \int {\rm d} \mathbf{s}\,
e^{-i \mathbf{q} \cdot \mathbf{s}} \mathbf{v(s)}$, 
as~\cite{saffman-76,diamant-09,IF-08}
\begin{equation}
v_i [\mathbf{q}]= G_{ij}^{\rm SD} [\mathbf{q}] f_j 
= \frac{1}{\eta q (q+\nu)} \left( \delta_{ij} - 
\frac{q_i q_j}{q^2}\right)f_j,
\end{equation}
where $G^{\rm SD}$ is the 2D film analog of the Oseen tensor.
In the above, the SD length is defined by 
$\nu^{-1} = \eta/2 \eta_{\rm s}$.

For over-damped dynamics, we can use the Einstein relation to relate 
the diffusion tensor $D_{\rm L}^{\rm c}$ to 
$G_{ij}^{\rm SD}$~\cite{diamant-09}. 
After converting $G_{ij}^{\rm SD}[\mathbf{q}]$ into real space, we 
obtain 
\begin{align}
D_{\rm L}^{\rm c}(s) & = k_{\rm B}T G_{xx}^{\rm SD}(s \hat{x}) 
\nonumber \\
& =\frac{k_{\rm B}T}{4\pi \eta} 
\left[\frac{\pi \mathbf{H}_1(\nu s)}{\nu s}-
\frac{\pi Y_1(\nu s)}{\nu s} 
- \frac{2}{(\nu s)^2} \right], 
\label{eqn:longD}
\end{align}
where $\mathbf{H}_1$ and $Y_1$ are Struve function and 
Bessel function of the second kind, respectively.  
At short distances $s \ll \nu^{-1}$, the asymptotic form of 
the above expression becomes  
\begin{equation}
D_{\rm L}^{\rm c}(s) \approx 
\frac{k_{\rm B}T}{4 \pi \eta} 
\left[ \ln\left(\frac{2}{\nu s}\right) - \gamma + \frac{1}{2}\right],
\label{eqn:longDsmallr}
\end{equation}
where $\gamma=0.5772 \cdots$ is Euler's constant. 
At large inter-particle separations $s \gg \nu^{-1}$, 
on the other hand, eq.~(\ref{eqn:longD}) reduces to  
\begin{equation}
D_{\rm L}^{\rm c}(s) \approx 
\frac{k_{\rm B}T}{2\pi\eta\nu s}=
\frac{k_{\rm B}T}{4\pi\eta_{\rm s} s},
\label{eqn:longDlarger}
\end{equation} 
showing the asymptotic $1/s$ decay which reflects the 3D nature 
of this limit.
Notice that eq.~(\ref{eqn:longDlarger}) depends only on the solvent
viscosity $\eta_{\rm s}$ but not on the film viscosity $\eta$
any more.

In fig.~\ref{fig:2pcorr}, we plot the measured longitudinal coupling 
diffusion coefficient $D_{\rm L}^{\rm c}$ as a function of 2D distance $s$.
In these simulations, we have worked with only single component 
 films with the same system sizes and number of particles 
as those used for the phase separation simulations.
We have also taken average over 20 independent trials. 
In the pure 2D case without any solvent, $D_{\rm L}^{\rm c}$ shows a 
logarithmic dependence on $s$. 
This is consistent with eq.~(\ref{eqn:longDsmallr}) obtained 
when the coupling between the film and solvent is very weak so 
that the film can be regarded almost as a pure 2D system. 
Using eq.~(\ref{eqn:longDsmallr}) as an approximate expression, we get
from the fitting as  
$k_{\rm B}T/4\pi\eta \approx 0.89\times10^{-2}$ and 
$\nu^{-1} \approx 20 $.
In an ideal case, the SD length should diverge due to the absence
of solvent.
The obtained finite value for $\nu^{-1}$ is roughly set by the
half of the system size in the simulation.

When we add solvent ($L_z=40$), the  $D_{\rm L}^{\rm c}$ is decreased 
and no longer behaves logarithmically.
In this case, we use the full expression eq.~(\ref{eqn:longD}) for the 
fitting, and obtained $k_{\rm B}T/4\pi\eta \approx 1.35\times 10^{-2}$ 
and $\nu^{-1} \approx 1$.
In the above two fits we have neglected the first two points as
they lie outside the range of validity, $s\gg1$, of 
eq.~(\ref{eqn:longD})~\cite{diamant-09}.
Since $\nu^{-1} \approx 1$ when the solvent is present, the data shown 
in fig.~\ref{fig:2pcorr} are in the crossover region,  
$s \gtrsim \nu^{-1}$, showing an approach towards the asymptotic $1/s$ 
dependence as in eq.~(\ref{eqn:longDlarger}).
Hence we conclude that the solvent brings in the 3D hydrodynamic 
property to the diffusion in films.
This is the reason for the 3D exponent $\alpha=1/3$ 
in the phase separation dynamics, and justifies that it is 
mainly driven by Brownian coagulation mechanism.

In our simulations the film and the solvent
have very similar viscosities.
This sets the SD length scale to be of the order of unity, which
is consistent with the value $\nu^{-1} \approx 1$ obtained from the
fitting.
As explained above, the fit also provides the 2D film viscosity as
$\eta\approx6$, and hence we obtain as $\eta_{\rm s} \approx 3$.
This value is in reasonable agreement with the value
$\eta_{\rm s} \approx 1$ calculated in ref.~\cite{pan-08} by using
the reverse Poiseuille flow method.
The reason for the slightly higher value of $\eta_{\rm s}$ in our 
simulations is that the tracer particles are of the same size as the
film particles.
This may lead to an underestimation of the correlated diffusion
coefficient.
In real biomembranes sandwiched by water, however, the value of the
SD length is much larger than the lipid size, and is in the order
of micron scale~\cite{saffman-76}.
Hence the 3D nature of hydrodynamics can be observed only for large enough
domains observed under optical microscopes~\cite{cicuta-07}.

\section{Discussion}

Several points merit further discussion. 
The growth exponents obtained from our simulations for critical mixtures
are the same for the off-critical case, namely $\alpha=1/2$ without 
solvent and $\alpha=1/3$ with solvent.  
A previous DPD study by Laradji and Kumar on phase separation dynamics of 
two-component membranes (both critical and off-critical cases) used a 
coarse-grained model for the membrane lipids~\cite{sunil-mohamed-05}. 
In their model, the self-assembly of the bilayer in the presence of  
solvent is naturally taken into account.
The exponent for the off-critical case $\alpha=1/3$ is the same as that 
obtained in our study, although they attributed this value to the LS 
mechanism.
For critical mixtures in the presence of solvent, they obtained a 
different value $\alpha=1/2$.
A suitable explanation for this exponent was not given in their paper.

As a related experimental work, the diffusion of tracer particles 
embedded in a soap film was recently reported~\cite{prasad-09}.
When the diameter of the tracer particles is close to the thickness 
of the soap film, the system shows a 2D behavior.
On the other hand, if the particle diameter is much smaller than the
soap film thickness, it executes a 3D motion.
On systematically increasing the soap film thickness, they identified a 
transition from 2D to 3D nature.
In this paper, we have demonstrated the analogue for a
2D liquid film-solvent system using DPD simulations.

In the SD theory, the bulk fluid is assumed to occupy an infinite 
space above and below the membrane. 
The situation is altered when the solvent and the membrane are confined 
between two solid walls~\cite{IF-08}. 
If the distance $H$ between the membrane and the wall is small enough, 
we are in a situation similar to that described by 
Evans and Sackmann~\cite{evans-88}.
The Oseen tensor $G^{\rm ES}$ in this case is defined through
the relation~\cite{shige-07},
\begin{equation}
v_i [\mathbf{q}]= G^{\rm ES}_{ij} [\mathbf{q}] f_j
= \frac{1}{\eta (q^2+\kappa^2)} 
\left( \delta_{ij} - \frac{q_i q_j}{q^2}\right) f_j,
\end{equation}
where the new length scale is defined as 
$\kappa^{-1}=\sqrt{\eta H/2\eta_{\rm s}}$.  
Notice that $\kappa^{-1}$ is the geometric mean of $\nu^{-1}$ and 
$H$~\cite{stone-98}.
Following the same procedure as in the previous section, the
longitudinal coupling diffusion coefficient can be obtained as 
\begin{equation}
D_{\rm L}^{\rm c}(s) = 
\frac{k_{\rm B}T}{2\pi\eta} \left[\frac{1}{(\kappa s)^2} 
- \frac{K_1(\kappa s)}{\kappa s} \right],
\label{ESdiff}
\end{equation}
where $K_1$ is modified Bessel function of the second kind.
At short distances $s \ll \kappa^{-1}$, we have 
\begin{equation}
D_{\rm L}^{\rm c}(s) \approx 
\frac{k_{\rm B}T}{4 \pi \eta} 
\left[ \ln\left(\frac{2}{\kappa s}\right) - \gamma + \frac{1}{2}\right],
\end{equation}
which is almost identical to eq.~(\ref{eqn:longDsmallr}) except
$\nu$ is replaced now by $\kappa$. 
At long distances $s \gg \kappa^{-1}$, on the other hand, we get  
\begin{equation}
D_{\rm L}^{\rm c}(s) \approx 
\frac{k_{\rm B}T}{2\pi\eta \kappa^2 s^2}=
\frac{k_{\rm B}T H}{4\pi\eta_{\rm s} s^2}, 
\end{equation} 
which exhibits a $1/s^2$ dependence. 
This is in contrast to eq.~(\ref{eqn:longDlarger}). 
Following the similar scaling argument as before, we predict 
that, in the presence of solid walls, the domain growth exponent 
should be $\alpha=1/4$ within the Brownian coagulation mechanism.
In biological systems, the above model with solid walls can be 
relevant because the cell membranes are strongly anchored to the 
underlying cytoskeleton, or are tightly adhered to other cells.

In summary, we have shown that the bulk fluid has a significant 
effect on the phase separation dynamics of  thin liquid films
through a simple quasi-2D simulation model.
We have demonstrated the change in the growth exponents
from 2D to 3D by the addition of bulk fluid.
This is further justified by the two-particle correlation studies, 
which showed that the longitudinal coupling diffusion coefficient 
changes from a logarithmic dependence for the pure 2D case to an 
algebraic one for the quasi-2D case.
Future directions include investigating the role of  out-of-plane 
fluctuations and the effect of boundary walls on the phase separation.

\acknowledgments

This work was supported by KAKENHI (Grant-in-Aid for Scientific
Research) on Priority Area ``Soft Matter Physics'' and Grant
No.\ 21540420 from the Ministry of Education, Culture, Sports, 
Science and Technology of Japan.



\begin{thebibliography}{99}
\bibitem{LS}
We note that the LS mechanism shows an exponent of $\alpha=1/3$ for both 
2D and 3D.
Thus our simulations are still in the early stage of the coarsening dynamics.
\end{thebibliography}
\end{document}